\documentclass[reprint,superscriptaddress,aps,prb,twocolumn,floatfix]{revtex4-2}
\usepackage[utf8]{inputenc}
\usepackage{textcomp} 
\usepackage{setspace}
\usepackage{amsmath}
\usepackage{breqn}
\usepackage{graphicx}
\usepackage{verbatim}

\usepackage{amsfonts}
\usepackage{amssymb}
\usepackage{xcolor}
\usepackage{soul}
\usepackage{tabularx}
\setcitestyle{super}
\usepackage{float}
\usepackage{mhchem}
\usepackage[normalem]{ulem}

\usepackage[colorlinks,linkcolor=blue,anchorcolor=blue,citecolor=blue,urlcolor=black]{hyperref}
\DeclareGraphicsExtensions{.pdf,.eps,.png,.jpg,.mps}

\begin{document}

\title{Multicolor interband solitons in microcombs}

\author{Qing-Xin Ji$^{1,\ast}$, Hanfei Hou$^{1,\ast}$, Jinhao Ge$^1$, Yan Yu$^1$, Maodong Gao$^1$, 
Warren Jin$^{2,3}$, Joel Guo$^2$, Lue Wu$^1$, Peng Liu$^1$, Avi Feshali$^3$, Mario Paniccia$^3$, John Bowers$^2$, Kerry Vahala$^{1,\dagger}$\\
$^1$T. J. Watson Laboratory of Applied Physics, California Institute of Technology, Pasadena, CA 91125, USA\\
$^2$ECE Department, University of California Santa Barbara, Santa Barbara, CA 93106, USA\\
$^3$Anello Photonics, Santa Clara, CA, USA\\
$^{\ast}$These authors contributed equally to this work.\\
$^{\dagger}$Corresponding authors: vahala@caltech.edu}

\maketitle

\noindent \textbf{In microcombs, solitons can drive non-soliton-forming modes to induce optical gain. Under specific conditions, a regenerative secondary temporal pulse coinciding in time and space with the exciting soliton pulse will form at a new spectral location. 
A mechanism involving Kerr-induced pulse interactions has been proposed theoretically, leading to multicolor solitons containing constituent phase-locked pulses. 
However, the occurrence of this phenomenon requires dispersion conditions that are not naturally satisfied in conventional optical microresonators. Here, we report the experimental observation of multicolor pulses from a single optical pump in a way that is closely related to the concept of multicolor solitons. The individual soliton pulses share the same repetition rate and could potentially be fully phase-locked. They are generated using interband coupling in a compound resonator.}

\medskip

\noindent \textbf{Introduction }
Dissipative solitons (DSs) in optical microresonators are self-reinforcing, localized wave packets generated through the double balance between propagation loss and nonlinear gain, as well as cavity dispersion and nonlinearity. In optical microresonators, various mechanisms for DS generation have been reported, including Kerr DS in $\chi^{(3)}$ medium \cite{herr2014temporal}, DS in optical parametric oscillators \cite{jankowski2018temporal}, Pockels DS in $\chi^{(2)}$ medium \cite{bruch2021pockels,skryabin2021sech}, and optomechanical DS \cite{zhang2021optomechanical}. 
Solitons can also drive non-soliton-forming modes to induce optical gain. A secondary temporal pulse coinciding in time and space with the exciting soliton pulse can form. 
The secondary pulse is not necessarily phase-locked with the exciting pulse, but is synchronized with its repetition rate. 
Stokes solitons are an example of this multicolor pulse behavior in which the gain is provided by the Raman interaction \cite{yang2016stokes}. 

Also, a new class of complex solitary wave has been theoretically proposed \cite{luo2016multicolor}, and referred to as multicolor solitons.  
The multicolor soliton landscape features dispersive waves originating from a primary soliton. These waves coherently pump another soliton (or several other solitons) via Kerr parametric gain at a different optical frequency (or several optical frequencies). 
The newly generated solitons coincide with the primary soliton in the temporal domain, and share the same group velocity with the primary soliton. 
The occurrence of this phenomenon requires that the dispersive waves are phase-matched, as well as that the local group velocity is matched between the primary soliton frequency and the dispersive wave's frequency. 
These combined dispersion requirements do not naturally exist in usual optical microresonators, and several engineered device structures have been numerically simulated to support such dispersion \cite{luo2016multicolor,moille2018phase-locked,silvestri2025theory}. Several related works use additional pump(s) to generate a secondary soliton (or non-solitonic microcomb) at a different color \cite{moille2022synthetic,gao2024multi,menyuk2025multi}. However, no experimental demonstrations of multicolor solitons with a single pump (as proposed in ref. \cite{luo2016multicolor}) have been implemented to our knowledge. 

Here, we report an experimental observation that is closely related to multicolor cavity solitons\cite{hou2025multicolor}. The required dispersion is achieved in a three-coupled-ring (3CR) microresonator and controlled via differential heater tuning of the rings. When the primary soliton is generated from a continuous wave (CW) pump laser, it can spontaneously trigger the formation of another soliton (hereafter referred to as secondary soliton) at a different carrier frequency and in a certain cavity-laser detuning regime. 
The secondary soliton is experimentally confirmed to be a femtosecond pulse that shares the same group velocity with the primary soliton. 
In contrast to the original multicolor soliton proposal \cite{luo2016multicolor}, the multicolor solitons observed here exist on distinct frequency bands (i.e., interband), and thus do not naturally share the same optical phase. 
However, feedback control of the pump laser is experimentally demonstrated to stabilize the relative optical phase between the two solitons.
The central carrier frequency difference of the two solitons at different colors can also be tuned electrically by differential heater tuning of the 3CR, ranging from 0.5 THz to 1.5 THz. The results enrich the soliton family. This approach to multicolor soliton generation can be used to extend the spectrum of the soliton. Also, the control and tuning capabilities of the multicolor solitons are potentially useful for high-coherence THz-wave generation.


\noindent{\bf Results}

\begin{figure*}
    \centering
    \includegraphics[width=\linewidth]{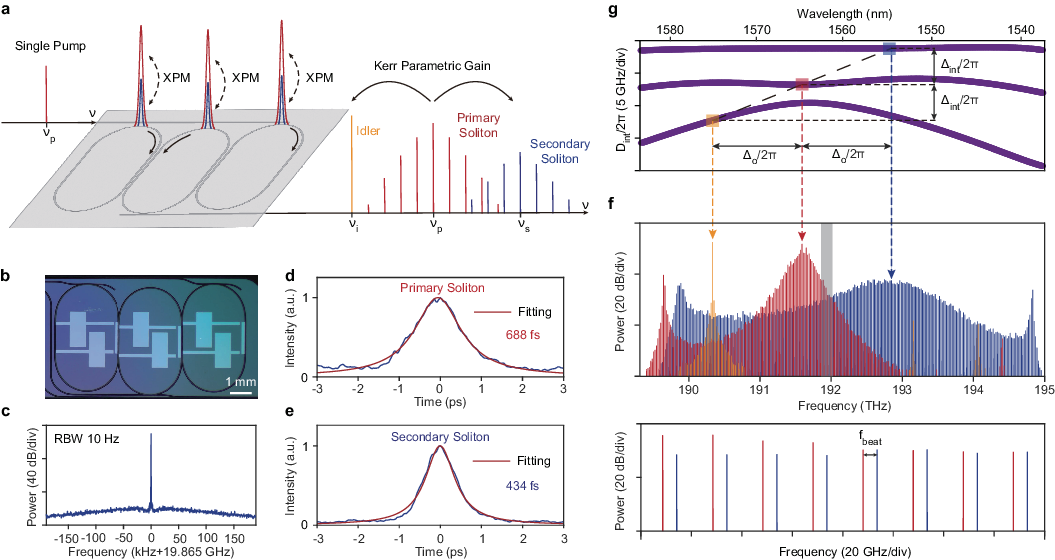}
    \caption{{\bf Generation of the multicolor interband solitons.} {\bf a,} 
    Conceptual illustration of multicolor interband soliton generation. The primary soliton is generated at the carrier frequency $\nu_{\mathrm{p}}$ from a single CW pump. The secondary soliton emerges at a different carrier frequency $\nu_{\rm s}$ (which coincides with the primary soliton in the time domain), accompanied by generation of a four-wave-mixing idler sideband at $\nu_{\mathrm{i}}$. The primary and secondary soliton trap each other to synchronize in the temporal domain via Kerr cross-phase modulation (XPM). 
    {\bf b,} Photograph of the three-coupled-ring device used in this study.
    {\bf c,} RF spectrum of detected pulse train (10 Hz resolution bandwidth). 
    {\bf d,e,} Measured autocorrelation traces (blue) and their Lorentzian fitting curves (red) for the primary and secondary solitons. Inferred full-width-half-maximum optical pulse widths are marked. 
    {\bf f,} Optical spectrum of multicolor interband solitons. The red, blue and orange spectral lines represent the primary soliton, secondary soliton and idler sideband, respectively. The lower panel is a zoom-in view of the gray shaded area in the upper panel, showing two different sets of comb lines corresponding to two solitons. The frequency spacing between adjacent comb lines is denoted by $f_{\mathrm{beat}}$. 
    {\bf g,} Dispersion spectrum for the generation of the multicolor interband solitons. The three-coupled-ring microresonator has three dispersion bands, and is pumped at a mode on the middle band (red). One mode on the upper band (blue) and one mode on the lower band (orange), together with the pumped mode, satisfy the phase-matching condition for parametric oscillation, as indicated by the black dashed line. The secondary soliton and idler sideband are generated near these two modes.}   
    \label{Fig1}
\end{figure*}

\begin{figure*}[t]
    \centering
    \includegraphics[width=\linewidth]{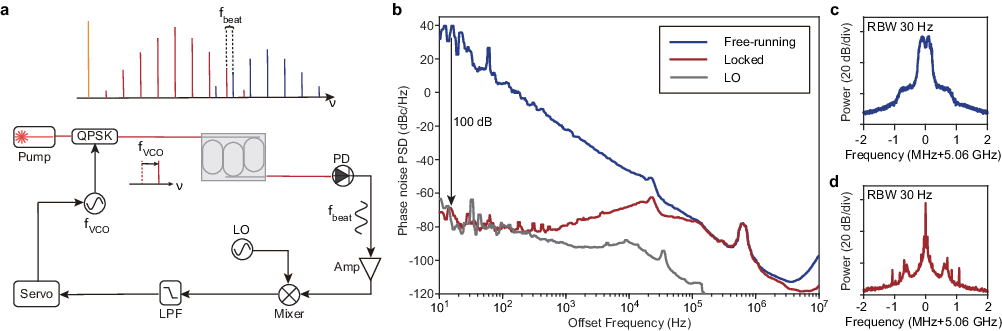}
    \caption{{\bf Phase stabilization of the multicolor interband solitons.} {\bf a,} Experimental setup. VCO, voltage-controlled oscillator, whose frequency is denoted by $f_{\rm VCO}$. QPSK, quadrature phase shift keying. PD, photodetector. Amp, electrical amplifier. LO, local oscillator. LPF, low pass filter. 
    {\bf b,} Single-sideband phase noise of free-running and locked inter-soliton beatnotes. Phase noise of local oscillator is also shown for comparison. {\bf c,d,}  RF spectra of free-running and locked inter-soliton beatnote tone.}
    \label{Fig2}
\end{figure*}

\begin{figure*}[t!]
    \centering
    \includegraphics[width=\linewidth]{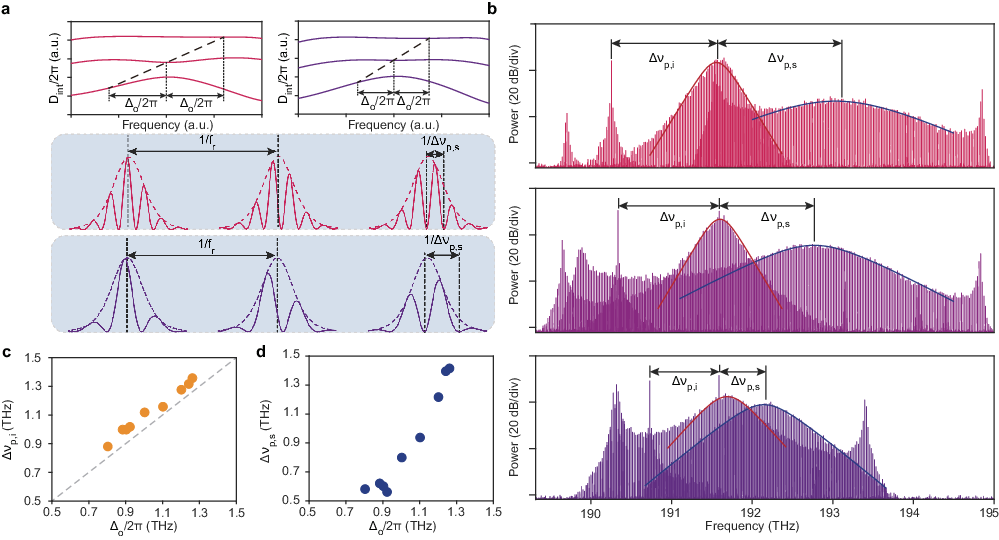}
    \caption{{\bf Spectral tuning of multicolor interband solitons.} {\bf a,} Conceptual illustration of the tuning mechanism. Cavity dispersion can be tuned by adjusting heater voltage parameters, leading to changes in frequency separation between primary and secondary solitons $\Delta \nu_{\rm p,s}$. In the time domain, interference between two solitons creates a pulse with THz-rate modulation. The modulation frequency $\Delta \nu_{\rm p,s}$ can be tuned by the heater tuning. {\bf b,}  Optical spectra with different $\Delta \nu_{\rm p,s}$ and $\Delta \nu_{\rm p,i}$ under different heater parameters. {\bf c,} $\Delta \nu_{\rm p,i}$ versus $\Delta_{\rm o}/2\pi$ under different heater parameters. {\bf d,} $\Delta \nu_{\rm p,s}$ versus $\Delta_{\rm o}/2\pi$ under different heater parameters.}
    \label{Fig3}
\end{figure*}

\noindent \textbf{Generation of multicolor interband solitons.} The device used in this work is a three-coupled-ring (3CR) microresonator (Fig. \ref{Fig1}a,b) \cite{yuan2023soliton,ji2025dispersive}. 
The scheme of multicolor interband solitons generation is illustrated in Fig. \ref{Fig1}a. A primary soliton (red) is first generated by pumping the microresonator 
with an amplified CW laser at $\nu_{\rm p}$. 
The primary soliton induces Kerr parametric gain and an effective potential well due to cross-phase-modulation (XPM) at its temporal location. 
The secondary soliton (blue) forms with a threshold behavior, through the double balance between parametric gain and cavity loss, as well as XPM and local anomalous dispersion (detailed in Methods).
An idler sideband (orange) is also formed as a result of the four-wave-mixing between the primary and secondary soliton, which cannot yield a soliton due to local normal dispersion in this case. 

The fact that the secondary soliton shares the same group velocity (repetition rate) is confirmed by the repetition rate measurement using a fast photodetector and an electrical spectral analyzer (Fig. \ref{Fig1}c). One single high-signal-to-noise ratio (SNR) tone is observed via photodetection. Here, to increase the signal stability for a higher SNR measurement, the primary soliton is disciplined to a stable microwave synthesizer \cite{weng2019spectral}, while similar results can also be observed without the microwave disciplining (Fig. \ref{Ext Datafig 1}). The temporal pulse nature of the two solitons is confirmed by auto-correlation measurement (Fig. \ref{Fig1}d,e) with the setup detailed in Methods. 
The primary soliton features a 688 fs full-width-half-maximum temporal duration, while that of the secondary soliton is 434 fs. 


Optical spectra of the generated multicolor solitons is presented in Fig. \ref{Fig1}f. The spectra are measured by collecting the output from the bus waveguide coupled to the middle ring (which serves an effective drop port), and measuring the output using a high resolution optical spectrum analyzer (APEX AP2083A, $\sim$10 MHz frequency resolution). In Fig. \ref{Fig1}f, the comb lines from the primary soliton are colored in red, while those of the secondary soliton (idler) are in blue (orange). 
As a modification to the multicolor solitons proposed in ref. \cite{luo2016multicolor}, the carrier-offset frequencies of two solitons are not necessarily the same. The lower panel of Fig. \ref{Fig1}f is a zoomed-in view of overlapping region between two solitons indicated by gray shade in the upper panel. 
Two sets of comb lines separated by a frequency of $f_{\rm beat}$ are observed. The results indicate no fixed phase relationship between the primary and secondary soliton is guaranteed. However, servo control of the pump laser is possible to force the phase locking, which will be detailed later.

\noindent \textbf{Dispersion condition for multicolor interband solitons generation.} The generation of multicolor solitons requires specific dispersion conditions. In this case, it is addressed by on-demand electrical tuning of the dispersion \cite{ji2024multimodality,ji2025dispersive}. 
The resonator dispersion spectrum that supports the optical spectrum in Fig. \ref{Fig1}f is shown in Fig. \ref{Fig1}g. 
Three hybrid mode families are formed, giving rise to three bands in the dispersion spectrum. 
The primary soliton is pumped at an anomalous dispersion window ($D_{\rm 2,p}/2\pi= 374$ kHz) on the middle band near 1565 nm. 

To generate the multicolor solitons (secondary soliton), it is first necessary to phase-match to the dispersive waves on the other bands. Here, interband phase-matching of parametric oscillation is achieved between the three dispersion bands. 
The parametric process involves two photons from the middle (pumped) band (frequency $\nu_{\rm p}$ close to the cavity resonance at frequency $\nu_{\rm p,c}$), and one photon from each of the upper and lower bands (whose frequencies $\nu_{\rm s}$, $\nu_{\rm i}$ are near the corresponding cavity resonance with frequencies$\nu_{\rm s,c}$ and $\nu_{\rm i,c}$), respectively, such that 
\begin{equation}
    \nu_{\rm s,c}+\nu_{\rm i,c}\approx2\nu_{\rm p,c}
    \label{eqn1}
\end{equation}
with the integrated dispersion at these modes $D_{\rm int,p}$, $D_{\rm int,s}$, $D_{\rm int,i}$ satisfying
\begin{equation}
    D_{\rm int,s}+ D_{\rm int,i}\approx 2D_{\rm int,p}
    \label{eqn0}
\end{equation}
for resonant excitation (phase matching). The phase-matched frequency is indicated by the black dashed line in Fig. \ref{Fig1}g. On the dispersion spectrum, the three modes that are phase-matched are equally-spaced both horizontally and vertically, with spacing $\Delta_{\rm o}/2\pi$, $\Delta_{\mathrm{int}}/2\pi$ respectively, as a result of eqn. \eqref{eqn1}\eqref{eqn0}.

The second requirement for the generation of the multicolor solitons is the group velocity matching of the primary and secondary solitons, so as to synchronize (and trap) the propagation of the two solitons along the resonator. Experimentally, the $FSR$s of the middle and upper band at $\nu_{\rm p,c}$ and $\nu_{\rm s,c}$ are measured to be near $19.86$ GHz with a slight difference of $\sim$ 1 MHz. 
The upper band simultaneously features local anomalous dispersion ($D_{\rm 2,s}/2\pi=39$ kHz), suitable for bright soliton mode locking. 
On the lower dispersion band, normal dispersion around the phase-matched mode does not support soliton formation, resulting in a sharp spectral peak.

\medskip

\noindent{\bf Servo phase-locking of the multicolor interband solitons.} Different from the multicolor solitons proposed in the ref. \cite{luo2016multicolor}, the multicolor interband solitons do not phase-lock as a result of the frequency offset between the dispersion bands. Here, we show that the phase locking can be achieved by servo control of the pump laser. Given that the repetition rates of the primary and secondary solitons are always the same, after the phase locking, the secondary soliton can be viewed as a coherent extension of the primary soliton. 

Experimental setup for $f_{\rm beat}$ locking is illustrated in Fig. \ref{Fig2}a. A fiber laser is frequency shifted by a quadratic phase shift keying (QPSK) driven by a voltage-controlled oscillator (VCO). The sideband from the QPSK (whose frequency is higher than the pump by the VCO frequency) is used to pump the multicolor interband solitons, followed by optical amplification. The microcomb output is amplified by an Erbium-doped fiber amplifier (EDFA) and directed to a fast photodetector, producing a beatnote signal $f_{\rm beat}$ at around 5 GHz. 
The beatnote is electrically amplified and mixed with a 5 GHz stable local oscillator (LO), generating an error signal. 
The servo output controls frequency of the VCO, which shifts the $f_{\rm beat}$ by controlling the pump line frequency $\nu_{\rm p}$ in Fig. \ref{Fig1}a. 
Phase noise of the locked $f_{\rm beat}$ is measured by a commercial phase noise analyzer (R\&S FSWP) as presented in Fig. \ref{Fig2}b in red, while that of the free-running $f_{\rm beat}$ is plotted in blue for comparison purpose. 
Phase noise of locked $f_{\rm beat}$ is 100 dB lower than the free-running case at 10 Hz frequency offset, and follows the LO phase noise (gray) at low offset frequencies ($<$ 100 Hz). 
RF spectra of the free-running and locked $f_{\rm beat}$ tone are presented in Fig. \ref{Fig2}c,d, respectively. Note that the $f_{\rm beat}$ locking is compatible with the simultaneous locking of $f_{\rm rep}$, for full phase stabilization between any of the comb lines of the multicolor solitons, which is detailed in Methods.

\medskip

\noindent{\bf Thermal tuning of the multicolor interband solitons. }
Electrical tuning of the frequency separation between the two solitons is demonstrated via differential temperature tuning between the three rings. 
The tuning scheme is illustrated in Fig. \ref{Fig3}a. Temperature of three rings in the cavity is independently controlled by electrical heaters. 
Resonator dispersion curve is tuned efficiently by adjusting heater voltage parameters \cite{ji2024multimodality,ji2025dispersive}. 
In Fig. \ref{Fig3}b, three optical spectra measured at different heater parameters are shown. 
Here, the primary soliton is fitted by the $\mathrm{sech}^2$ envelope, while fitting details of the secondary soliton are detailed in Methods. 

When the dispersion is modified, the phase-matched frequency of the Cherenkov radiation is actively tuned (while the pump frequency is fixed). 
The idler sideband is expected to emerge at the phase-matched mode, and the frequency separation between the pump soliton spectral center and the idler sideband $\Delta \nu_{\rm p,i}$ is predicted to be $\Delta_{\rm o}/2\pi$. 
In Fig. \ref{Fig3}c, measured values of $\Delta \nu_{\rm p,i}$ versus $\Delta_{\rm o}/2\pi$ (derived from dispersion measurements) are plotted. 
As the central wavelength of the secondary soliton mostly follows the phase-matched wavelength, the frequency separation between the pump and the secondary soliton spectral center $\Delta \nu_{\rm p,s}$ 
is also actively tuned. 
Measured values of $\Delta \nu_{\rm p,s}$ are presented in Fig. \ref{Fig3}d. 

\medskip

\noindent \textbf{Discussion and conclusion. }
A potential application of multicolor interband solitons is terahertz wave generation. The frequency separation between two solitons $\Delta \nu_{\rm p,s}$ falls within THz range, and in the time domain the interference between two solitons creates a THz modulation in optical intensity. In the time domain, after averaging out optical frequency oscillations, temporal optical intensity distribution features pulses with THz-band carrier. Pulse repetition rate is still $f_{\rm rep}$ and carrier frequency is $\Delta \nu_{\rm p,s}$, which is tunable. The two boxes with different colors on the right of Fig. \ref{Fig3}a show temporal pulse trains with different carrier frequencies under different heater parameters. By converting the optical wave into terahertz waves using photoconductive process \cite{fattinger1989terahertz} or optical rectification \cite{yeh2007generation}, a THz-band frequency comb  is generated, where the central frequency is $\Delta \nu_{\rm p,s}\sim 0.5-1.5$ THz, and the repetition rate is $f_{\rm rep}\approx20$ GHz. 


In conclusion, a new type of soliton closely related to the phenomenon theoretically predicted in the reference \cite{luo2016multicolor} is experimentally demonstrated. A secondary soliton is generated from a primary soliton via Kerr parametric gain and trapped by the potential well created by cross phase modulation from the primary soliton. The two solitons have different central frequency, but coincide in time and share common repetition rate. The two-soliton microcomb can be fully referenced to RF sources and has good tunability through dispersion tuning of the three-coupled-ring microresonator. This comb system is potentially useful as a chip-based terahertz comb source. The new physics also enriches understanding of cavity nonlinear soliton dynamics and points out a possible method of soliton generation and spectrum extension.

\bibliography{ref}

\providecommand{\noopsort}[1]{}\providecommand{\singleletter}[1]{#1}%
\begin{thebibliography}{10}
\expandafter\ifx\csname url\endcsname\relax
  \def\url#1{\texttt{#1}}\fi
\expandafter\ifx\csname urlprefix\endcsname\relax\def\urlprefix{URL }\fi
\providecommand{\bibinfo}[2]{#2}
\providecommand{\eprint}[2][]{\url{#2}}

\bibitem{herr2014temporal}
\bibinfo{author}{Herr, T.} \emph{et~al.}
\newblock \bibinfo{title}{Temporal solitons in optical microresonators}.
\newblock \emph{\bibinfo{journal}{Nat. Photon.}} \textbf{\bibinfo{volume}{8}}, \bibinfo{pages}{145--152} (\bibinfo{year}{2014}).

\bibitem{jankowski2018temporal}
\bibinfo{author}{Jankowski, M.} \emph{et~al.}
\newblock \bibinfo{title}{Temporal simultons in optical parametric oscillators}.
\newblock \emph{\bibinfo{journal}{Phys. Rev. Lett.}} \textbf{\bibinfo{volume}{120}}, \bibinfo{pages}{053904} (\bibinfo{year}{2018}).

\bibitem{bruch2021pockels}
\bibinfo{author}{Bruch, A.~W.} \emph{et~al.}
\newblock \bibinfo{title}{Pockels soliton microcomb}.
\newblock \emph{\bibinfo{journal}{Nat. Photon.}} \textbf{\bibinfo{volume}{15}}, \bibinfo{pages}{21--27} (\bibinfo{year}{2021}).

\bibitem{skryabin2021sech}
\bibinfo{author}{Skryabin, D.~V.}
\newblock \bibinfo{title}{Sech-squared pockels solitons in the microresonator parametric down-conversion}.
\newblock \emph{\bibinfo{journal}{Opt. Express}} \textbf{\bibinfo{volume}{29}}, \bibinfo{pages}{28521--28529} (\bibinfo{year}{2021}).

\bibitem{zhang2021optomechanical}
\bibinfo{author}{Zhang, J.} \emph{et~al.}
\newblock \bibinfo{title}{Optomechanical dissipative solitons}.
\newblock \emph{\bibinfo{journal}{Nature}} \textbf{\bibinfo{volume}{600}}, \bibinfo{pages}{75--80} (\bibinfo{year}{2021}).

\bibitem{yang2016stokes}
\bibinfo{author}{Yang, Q.-F.}, \bibinfo{author}{Yi, X.}, \bibinfo{author}{Yang, K.~Y.} \& \bibinfo{author}{Vahala, K.}
\newblock \bibinfo{title}{Stokes solitons in optical microcavities}.
\newblock \emph{\bibinfo{journal}{Nat. Phys.}} \textbf{\bibinfo{volume}{13}}, \bibinfo{pages}{53--57} (\bibinfo{year}{2017}).

\bibitem{luo2016multicolor}
\bibinfo{author}{Luo, R.}, \bibinfo{author}{Liang, H.} \& \bibinfo{author}{Lin, Q.}
\newblock \bibinfo{title}{Multicolor cavity soliton}.
\newblock \emph{\bibinfo{journal}{Opt. Express}} \textbf{\bibinfo{volume}{24}}, \bibinfo{pages}{16777--16787} (\bibinfo{year}{2016}).

\bibitem{moille2018phase-locked}
\bibinfo{author}{Moille, G.}, \bibinfo{author}{Li, Q.}, \bibinfo{author}{Kim, S.}, \bibinfo{author}{Westly, D.} \& \bibinfo{author}{Srinivasan, K.}
\newblock \bibinfo{title}{Phased-locked two-color single soliton microcombs in dispersion-engineered \ce{Si3N4} resonators}.
\newblock \emph{\bibinfo{journal}{Opt. Lett.}} \textbf{\bibinfo{volume}{43}}, \bibinfo{pages}{2772--2775} (\bibinfo{year}{2018}).

\bibitem{silvestri2025theory}
\bibinfo{author}{Silvestri, C.}, \bibinfo{author}{Widjaja, J.}, \bibinfo{author}{Lin, A.}, \bibinfo{author}{de~Sterke, C.~M.} \& \bibinfo{author}{Runge, A.~F.}
\newblock \bibinfo{title}{Theory of multicolor soliton microcombs}.
\newblock \emph{\bibinfo{journal}{Opt. Lett.}} \textbf{\bibinfo{volume}{50}}, \bibinfo{pages}{2073--2076} (\bibinfo{year}{2025}).

\bibitem{moille2022synthetic}
\bibinfo{author}{Moille, G.}, \bibinfo{author}{Menyuk, C.}, \bibinfo{author}{Chembo, Y.~K.}, \bibinfo{author}{Dutt, A.} \& \bibinfo{author}{Srinivasan, K.}
\newblock \bibinfo{title}{Synthetic frequency lattices from an integrated dispersive multi-color soliton}.
\newblock \emph{\bibinfo{journal}{arXiv preprint arXiv:2210.09036}}  (\bibinfo{year}{2022}).

\bibitem{gao2024multi}
\bibinfo{author}{Gao, M.} \emph{et~al.}
\newblock \bibinfo{title}{Multi-color solitons in coupled-ring microresonators}.
\newblock In \emph{\bibinfo{booktitle}{CLEO: Science and Innovations}}, \bibinfo{pages}{SM3G--1} (\bibinfo{organization}{Optica Publishing Group}, \bibinfo{year}{2024}).

\bibitem{menyuk2025multi}
\bibinfo{author}{Menyuk, C.~R.}, \bibinfo{author}{Shandilya, P.}, \bibinfo{author}{Courtright, L.}, \bibinfo{author}{Moille, G.} \& \bibinfo{author}{Srinivasan, K.}
\newblock \bibinfo{title}{Multi-color solitons and frequency combs in microresonators}.
\newblock \emph{\bibinfo{journal}{Opt. Express}} \textbf{\bibinfo{volume}{33}}, \bibinfo{pages}{21824--21835} (\bibinfo{year}{2025}).

\bibitem{hou2025multicolor}
\bibinfo{author}{Hou, H.} \emph{et~al.}
\newblock \bibinfo{title}{Multicolor interband solitons in microcombs}.
\newblock In \emph{\bibinfo{booktitle}{CLEO: Fundamental Science}} (\bibinfo{organization}{Optica Publishing Group}, \bibinfo{year}{2025}).

\bibitem{yuan2023soliton}
\bibinfo{author}{Yuan, Z.} \emph{et~al.}
\newblock \bibinfo{title}{Soliton pulse pairs at multiple colours in normal dispersion microresonators}.
\newblock \emph{\bibinfo{journal}{Nature Photonics}} \textbf{\bibinfo{volume}{17}}, \bibinfo{pages}{977--983} (\bibinfo{year}{2023}).

\bibitem{ji2025dispersive}
\bibinfo{author}{Ji, Q.-X.} \emph{et~al.}
\newblock \bibinfo{title}{Dispersive-wave-agile optical frequency division}.
\newblock \emph{\bibinfo{journal}{Nature Photonics}} \bibinfo{pages}{1--6} (\bibinfo{year}{2025}).

\bibitem{weng2019spectral}
\bibinfo{author}{Weng, W.} \emph{et~al.}
\newblock \bibinfo{title}{Spectral purification of microwave signals with disciplined dissipative kerr solitons}.
\newblock \emph{\bibinfo{journal}{Physical review letters}} \textbf{\bibinfo{volume}{122}}, \bibinfo{pages}{013902} (\bibinfo{year}{2019}).

\bibitem{ji2024multimodality}
\bibinfo{author}{Ji, Q.-X.} \emph{et~al.}
\newblock \bibinfo{title}{Multimodality integrated microresonators using the moir{\'e} speedup effect}.
\newblock \emph{\bibinfo{journal}{Science}} \textbf{\bibinfo{volume}{383}}, \bibinfo{pages}{1080--1083} (\bibinfo{year}{2024}).

\bibitem{fattinger1989terahertz}
\bibinfo{author}{Fattinger, C.} \& \bibinfo{author}{Grischkowsky, D.}
\newblock \bibinfo{title}{Terahertz beams}.
\newblock \emph{\bibinfo{journal}{Applied Physics Letters}} \textbf{\bibinfo{volume}{54}}, \bibinfo{pages}{490--492} (\bibinfo{year}{1989}).

\bibitem{yeh2007generation}
\bibinfo{author}{Yeh, K.-L.}, \bibinfo{author}{Hoffmann, M.}, \bibinfo{author}{Hebling, J.} \& \bibinfo{author}{Nelson, K.~A.}
\newblock \bibinfo{title}{Generation of 10$\mu$j ultrashort terahertz pulses by optical rectification}.
\newblock \emph{\bibinfo{journal}{Applied Physics Letters}} \textbf{\bibinfo{volume}{90}} (\bibinfo{year}{2007}).

\bibitem{lugiato1987spatial}
\bibinfo{author}{Lugiato, L.~A.} \& \bibinfo{author}{Lefever, R.}
\newblock \bibinfo{title}{Spatial dissipative structures in passive optical systems}.
\newblock \emph{\bibinfo{journal}{Phys. Rev. Lett.}} \textbf{\bibinfo{volume}{58}}, \bibinfo{pages}{2209} (\bibinfo{year}{1987}).

\end{thebibliography}

\medskip

\begin{figure*}
    \centering
    \includegraphics[width=\linewidth]{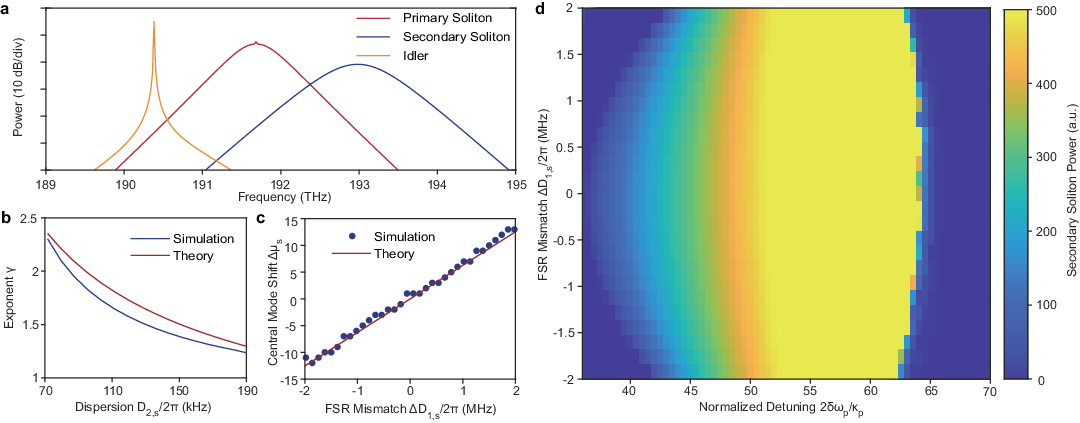}
    \caption{{\bf Simulation results.} {\bf a,} Simulated optical spectrum. {\bf b,} Comparison of theoretical prediction and simulation result of secondary soliton pulse profile exponent $\gamma$ versus second-order dispersion parameter $D_{\rm 2,s}$. {\bf c,} Comparison of theoretical prediction and simulation result of secondary soliton central mode shift $\Delta\mu_{\rm s}$ versus FSR mismatch $\Delta D_{\rm 1,s}/2\pi$. {\bf d,} Existence range of secondary soliton. Secondary soliton powers at different pump detunings $\delta \omega_{\rm p}$ and FSR mismatches are plotted.} 
    \label{Fig5}
\end{figure*} 

\begin{figure*}
    \centering
    \includegraphics[width=\linewidth]{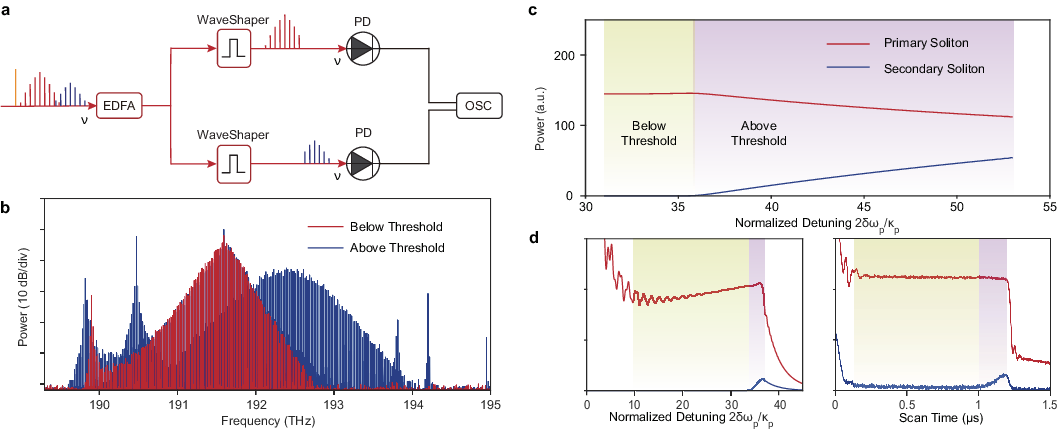}
    \caption{{\bf Threshold behavior.} {\bf a,} Experimental setup for soliton step measurement. EDFA, erbium-doped fiber amplifier. PD, photodetector. OSC, oscilloscope. {\bf b,} Spectra below and above threshold measured under the same experimental conditions. {\bf c,} Simulation result of primary and secondary soliton power versus normalized pump detuning $2\delta\omega_{\rm p}/\kappa_{\rm p}$ when the detuning is slowly ramped. When $2\delta\omega_{\rm p}/\kappa_{\rm p}<35.7$, secondary soliton power is close to zero (below threshold). When $2\delta\omega_{\rm p}/\kappa_{\rm p}>35.7$, secondary soliton power begins to increase, accompanied by a decrease in primary soliton power. Regions below (above) the threshold detuning is shaded in yellow (purple). {\bf d,} Simulated (left) and experimentally measured (right) soliton steps for primary and secondary solitons when pump detuning is scanned quickly. After the formation of the primary soliton, the secondary soliton does not emerge until a certain threshold detuning is reached. }
    \label{Ext Datafig 3}
\end{figure*}

\noindent {\bf Methods }\\
\noindent {\bf Theory of multicolor interband solitons generation.}
Here, the coupled rings are effectively replaced with a single cavity, and the three supermode families in Fig. \ref{Fig1}g are viewed as independent transverse mode families. 
The assumption is validated in the Supplementary Information. 
The soliton dynamics in presence of parametric interaction is governed by coupled Lugiato-Lefever equations \cite{lugiato1987spatial} with four-wave-mixing terms, which read
 \begin{equation}
\begin{split}
    \frac{\partial}{\partial T}E_{\rm p}
    &=-(\frac{\kappa_{\rm p}}{2}+i\delta \omega_{\rm p})E_{\rm p}+i\frac{D_{\rm 2,p}}{2}\frac{\partial^2}{\partial \phi^2}E_{\rm p}+2ig_{\rm FWM}^{\ast}E_{\rm s}E_{\rm i}E_{\rm p}^{\ast}\\
    &+i(g_0|E_{\rm p}|^2+2g_{\rm XPM}|E_{\rm s}|^2+2g_{\rm XPM}|E_{\rm i}|^2)E_{\rm p}+F, 
\end{split} 
\label{eqn22}
\end{equation}
\begin{equation}
\begin{split}
    \frac{\partial}{\partial T}E_{\rm s}
    &=-(\frac{\kappa_{\rm s}}{2}+i\delta \omega_{\rm s})E_{\rm s}-\Delta D_{\rm 1,s}\frac{\partial}{\partial \phi}E_{\rm s}+i\frac{D_{\rm 2,s}}{2}\frac{\partial^2}{\partial \phi^2}E_{\rm s}\\
    &+i(g_0|E_{\rm s}|^2+2g_{\rm XPM}|E_{\rm p}|^2+2g_{\rm XPM}|E_{\rm i}|^2)E_{\rm s}\\
    &+ig_{\rm FWM}E_{\rm p}^2E_{\rm i}^{\ast}, 
\end{split} 
\label{eqn23}
\end{equation}
\begin{equation}
\begin{split}
    \frac{\partial}{\partial T}E_{\rm i}
    &=-(\frac{\kappa_{\rm i}}{2}+i\delta \omega_{\rm i})E_{\rm i}-\Delta D_{\rm 1,i}\frac{\partial}{\partial \phi}E_{\rm i}+i\frac{D_{\rm 2,i}}{2}\frac{\partial^2}{\partial \phi^2}E_{\rm i}\\
    &+i(g_0|E_{\rm i}|^2+2g_{\rm XPM}|E_{\rm p}|^2+2g_{\rm XPM}|E_{\rm s}|^2)E_{\rm i}\\
    &+ig_{\rm FWM}E_{\rm p}^2E_{\rm s}^{\ast}. 
\end{split} 
\label{eqn24}
\end{equation}
Here, the slow-varying electrical fields $E_{\rm k}$ ($\rm k=p,s,i$ for primary soliton, secondary soliton and idler sideband respectively, * denotes complex conjugate) are defined in the co-rotating frame of the primary soliton and normalized to photon number. The carrier angular frequencies are denoted by $\omega_{\rm k}$ ($\rm k=p,s,i$), respectively.
$\omega_{\rm p}$ equals to pump angular frequency. 
The choice of $\omega_{\rm s}$ and $\omega_{\rm i}$ is not deterministic, but to eliminate the phase factor in the FWM terms, it is forced that
\begin{equation}
    \omega_{\rm s}+\omega_{\rm i}=2\omega_{\rm p}. 
\end{equation}
We further define detuning $\delta \omega_{\rm k}=\omega_{\rm k,c}-\omega_{\rm k}$ ($\omega_{\rm k,c}$ is the corresponding cavity mode angular frequency, $\rm k=p,s,i$), where
\begin{equation}
    \delta \omega_{\rm s}+\delta \omega_{\rm i}-2\delta \omega_{\rm p}
    \equiv\omega_{\rm s,c}+\omega_{\rm i,c}-2\omega_{\rm p,c}
    \equiv \mathrm{Constant}. 
\end{equation}
Since the detunings need to be small for resonant excitation, a requirement for mode frequencies eqn. \eqref{eqn1} arises.
Furthermore, $\kappa_{\rm k}$ ($\rm k=p,s,i$) is cavity loss rate, $\Delta D_{\rm 1,s}\equiv D_{\rm 1,s}-D_{\rm 1,p}$, $\Delta D_{\rm 1,i}\equiv D_{\rm 1,i}-D_{\rm 1,p}$, $D_{\rm 1,k}$, $D_{\rm 2,k}$ are first- and second- order cavity dispersion parameters, $g_{0}$, $g_{\rm XPM}$, $g_{\rm FWM}$ are effective nonlinear self-phase-modulation, cross-phase-modulation and four-wave-mixing coefficients respectively (for definition see Supplementary Information), and $F=\sqrt{\kappa_{\rm ext,p} P_{\rm in}/\hbar\omega_{\rm p}}$ is the pump term, where $\kappa_{\rm ext,p}$ is the external coupling rate and $P_{\rm in}$ is the on-chip input power.

\noindent \textbf{Analytical analysis. }
Several approximations are made to derive the analytical solution of eqn. (\ref{eqn22})(\ref{eqn23})(\ref{eqn24}). We focus on near-threshold behavior where the power of the secondary soliton and idler sideband is much lower than the primary soliton, that $\left|E_{\rm s}\right|$,$\left|E_{\rm i}\right| \ll \left|E_{\rm p}\right|$. 
The primary soliton takes the unperturbed soliton form
\begin{equation}
    E_{\rm p}=A_{\rm p} \mathrm{sech}{(B\phi)}, 
    \label{eqn5}
\end{equation}
The dynamics of $E_{\rm s}$ and $E_{\rm i}$, with this $E_{\rm p}$ expression inserted, yields Schr$\rm{\ddot{o}}$dinger-type equations in a $\mathrm{sech^2}$ potential well with parametric gain terms. The idler sideband is approximated as a continuous wave, while the secondary soliton exhibits ground-state solution as follows:
\begin{equation}
    E_{\rm s}= A_{\rm s}\mathrm{sech}^\gamma{(B\phi)}e^{-i\Delta\mu_{\rm s}\phi}, 
    \label{eqn6}
\end{equation}
\begin{equation}
    E_{\rm i}=A_{\rm i}. 
    \label{eqn7}
\end{equation}
The linear phase factor in $E_{\rm s}$ results from $FSR$ mismatch of the primary and secondary soliton forming mode families. $\Delta\mu_{\rm s}$ denotes a shift in secondary soliton central mode from mode $\omega_{\rm s,c}$. For $\gamma$ and $\Delta\mu_{\rm s}$ it is derived (detailed in the Supplementary Information),
\begin{equation}
    \gamma (1+\gamma)=\frac{4g_{\rm XPM}}{g_{0}}\frac{D_{\rm 2,p}}{D_{\rm 2,s}}. 
        \label{eqn16}
\end{equation}
\begin{equation}
\Delta\mu_{\rm s}=\frac{\Delta D_{\rm 1,s}}{D_{\rm 2,s}}. 
\label{eqn2}
\end{equation}
Eqn. \eqref{eqn2} indicates that the central mode of secondary soliton is shifted to where the $FSR$ of the soliton forming mode aligns with the primary soliton.

Furthermore, a threshold behaviour is predicted. $E_{\rm s}$ and $E_{\rm i}$ compose a coupled linear system, where either exponential growth or decay can occur. The secondary soliton forms under exponential growth, when parametric gain overcomes cavity loss. By taking the inner product of both sides of eqn. (\ref{eqn23})(\ref{eqn24}) with their respective eigenfunctions (\ref{eqn6})(\ref{eqn7}), the equations reduce to a linear set of ordinary differential equations governing the evolution of $E_{\rm s}$ and $E_{\rm i}$ amplitudes, and the threshold condition is readily obtained. Setting $\Delta D_{\rm 1,s}=0$ for simplicity of expression, its threshold condition is calculated to be 
\begin{equation}
\begin{split}
&\frac{\kappa_{\rm s} \kappa_{\rm i}}{4}+(\frac{\delta \omega_{\rm s} +\delta\omega_{\rm i}}{2}-\frac{g_{\rm XPM}\sqrt{2D_{\rm 2,p}\delta\omega_{\rm p}}}{\pi g_{0}}-\gamma^{2}\frac{D_{\rm 2,s}}{2D_{\rm 2,p}}\delta\omega_{\rm p})^2\\
&-\frac{2|g_{\rm FWM}|^2\delta\omega_{\rm p}^2}{\pi g_{0}^2}\sqrt{\frac{D_{\rm 2,p}}{2\delta\omega_{\rm p}}}\frac{\Pi(\gamma+2)^2}{\Pi(2\gamma)}=0. 
 \label{eqn21}
 \end{split}
\end{equation}
where $\Pi(t)\equiv \int_{-\infty}^\infty \mathrm{sech}^tx dx$.

\noindent {\bf Numerical Simulation. } 
In confirmation on that the proposed mechanism enables multicolor interband solitons, numerical simulations are performed  based on full coupled LLEs eqn. \eqref{eqn22}-\eqref{eqn24} using split-step Fourier transform method. For each dispersion band, 1024 modes are involved in the model.  

In the simulation, 
the system is seeded by a single primary soliton. 
The results are summarized in Fig. \ref{Fig5}. Simulated spectrum in Fig. \ref{Fig5}a displays good similarity to experimental data in Fig. \ref{Fig1}. The conclusions drawn from analytical model are also validated. To verify  eqn. \eqref{eqn16}, dispersion parameter $D_{\rm 2,s}$ is tuned and exponent $\gamma$ is determined by spectrum fitting at each $D_{\rm 2,s}$. The analytical and numerical results are consistent (Fig. \ref{Fig5}b). For eqn. \eqref{eqn2}, the central mode shift of the secondary soliton $\Delta \mu_{\rm s}$ obtained from simulation and eqn. \eqref{eqn2} are plotted together at different FSR mismatches $\Delta D_{\rm 1,s}$ in Fig. \ref{Fig5}c, also showing good agreement. 

Simulation parameters are listed as below. For Fig. \ref{Fig5}a, $\omega_{\rm p}=2\pi\times191.68$ THz, $Q_{\rm int}=75\times10^6$, $Q_{\rm ext}=200\times10^6$, $\delta\omega_{\rm p}=22.5\kappa_{\rm p}$, $\delta\omega_{\rm s}=\delta\omega_{\rm i}=12.5\kappa_{\rm p}$, $\Delta D_{\rm 1,s}=0$, $\Delta D_{\rm 1,i}=2\pi\times31.8$ MHz, $D_{\rm 2,p}=2\pi\times353$ kHz, $D_{\rm 2,s}=2\pi\times159$ kHz, $D_{\rm 2,i}=-2\pi\times154$ kHz, $g_0=0.0272$ $\rm Hz$, $g_{\rm XPM}=0.0109$ $\rm Hz$, $g_{\rm FWM}=0.0109$ $\rm Hz$, $P_{\rm in}=300$ mW. Loss rates are derived by $\kappa_{\rm ext,p}=\omega_{\rm p}/Q_{\rm ext}$, $\kappa_{\rm k}=\kappa_{\rm int}+\kappa_{\rm ext}=\omega_{\rm p}/Q_{\rm int}+\omega_{\rm p}/Q_{\rm ext}$ ($\rm k=p,s,i$). For Fig. \ref{Fig5}b, $\Delta D_{\rm 1,s}$ is fixed at 0. For Fig. \ref{Fig5}c,d, $D_{\rm 2,s}$ is fixed at $2\pi\times159$ kHz.

\noindent {\bf Threshold behavior of the secondary soliton generation.} Threshold behavior is observed both experimentally and numerically, which is typical for parametric processes and predicted by the theory. When the pump laser scans across the mode from blue-detuned regime to red-detuned regime, a single primary soliton is generated at first, and the secondary soliton and idler sideband emerge when pump detuning reaches a certain threshold. Optical spectrum below and above threshold detuning are shown in Fig. \ref{Ext Datafig 3}b. 

Measured and simulated soliton steps of primary and secondary solitons are also shown separately in Fig. \ref{Ext Datafig 3}d. The measurement setup is detailed in Fig. \ref{Ext Datafig 3}a. The comb output is amplified by an erbium-doped fiber amplifier (EDFA) and equally split into two routes. Each route is directed to an optical waveshaper to filter out primary/secondary soliton only, and then detected by a photodetector (PD). The PD signals are received by an oscilloscope to monitor soliton power. Analogous to simulation result, the measurement also verifies the sequenced generation of primary and secondary solitons, indicating the existence of threshold detuning. In the numerical model, when pump detuning is slowly ramped, existence of threshold is observed explicitly (Fig. \ref{Ext Datafig 3}c). Normalized threshold detuning $2\delta\omega_{\rm p}/\kappa_{\rm p}$ calculated from eqn. \eqref{eqn21} with simulation parameters is 33.8, close to simulation value 35.7.

Furthermore, threshold behavior in presence of FSR mismatch $\Delta D_{\rm 1,s}/2\pi$ is studied. Comb spectra are simulated under different FSR mismatches and pump detunings $\delta \omega_{\rm p}$, and secondary soliton powers with respect to these parameters are plotted in Fig. \ref{Fig5}d. Secondary soliton existence range is nearly symmetric with respect to $\Delta D_{\rm 1,s}/2\pi=0$, and threshold pump detuning increases with FSR mismatch. The primary and secondary solitons simultaneously vanish when pump detuning exceeds the primary soliton existence limit.

\noindent\textbf{Experimental details.} In the autocorrelation measurement, the comb output from the cavity is firstly amplified to 70 mW by an erbium-doped fiber amplifier (EDFA), and then directed to a waveshaper. The waveshaper is programmed as a band-pass filter that filters out either the primary or the secondary soliton, and in its passband, a quadratic dispersion is applied to compensate fiber dispersion. After filtering, the comb is again amplified to 300 mW by a second EDFA before sent into an autocorrelator. The data in Fig. \ref{Fig1}d,e is measured when dispersion compensation is optimized so that the pulses display the smallest temporal widths.

Full phase stabilization is achieved by simultaneous locking of $f_{\rm rep}$ (by servo locking) and $f_{\rm beat}$ (by disciplining to a stable microwave synthesizer). It can be characterized by another inter-soliton beatnote with frequency $f_{\rm rep}-f_{\rm beat}$. The noise of this beatnote will be significantly suppressed only when $f_{\rm rep}$ and $f_{\rm beat}$ are simultaneously stabilized. RF spectrum and phase noise data for the $f_{\rm rep}-f_{\rm beat}$ beatnote is plotted in Fig. \ref{Ext Datafig 4}. The phase noise of locked $f_{\rm rep}-f_{\rm beat}$ beatnote closely follows that of locked $f_{\rm beat}$ beatnote, while the noise of locked $f_{\rm rep}$ is below this level, indicating successful full phase stabilization. The two solitons form a coherent set of frequency comb with broader spectral range.

In Fig. \ref{Fig3}c and Fig. \ref{Fig5}b, consistent with the proposed theory, the secondary soliton spectra are fitted by an envelope with form
\begin{equation}
P(\nu)=\left|\mathrm{FT}\{a~\mathrm{sech}^\gamma (bt)\}(\nu-\nu_{\rm s})\right|^2, 
\end{equation}
where FT denotes Fourier transform, $P$ is power, $\nu$ is frequency, $t$ is time, and $a$, $b$, $\gamma$, $\nu_{\rm s}$ are fitting parameters.

\medskip

\noindent {\bf Funding}\\
This work was supported by AFOSR (FA9550-23-1-0587).\\
\noindent {\bf Acknowledgments} \\
The authors thank Curtis Menyuk at Maryland University, Bumki Min at KAIST, and Qi-Fan Yang at Peking University for discussions. \\
\noindent{\bf Author contributions} \\
Q.-X.J. and H.H. performed the measurements with assistance from J.Ge, M.G. and P.L. 
H.H. Y.Y. and Q-X.J. performed analytical modeling and numerical simulation. W.J., J.Guo and L.W. fabricated the device with assistance from A.F. and M.P.
All the authors contributed to the writing of the manuscript. 
J.B. and K.V. supervised the project. \\
\noindent{\bf Competing interests}\\
The authors declare no competing financial interests.\\
\noindent{\bf Data and materials availability} \\
The data that support the plots within this paper and the other findings of this study are available from the corresponding author upon reasonable request. 

\begin{figure}
    \centering
    \includegraphics[width=\linewidth]{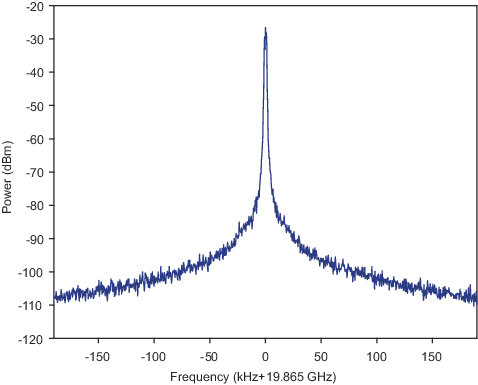}
    \caption{{\bf Free-running repetition rate tone without filtering.}}
    \label{Ext Datafig 1}
\end{figure}
\begin{figure*}
    \centering
    \includegraphics[width=\linewidth]{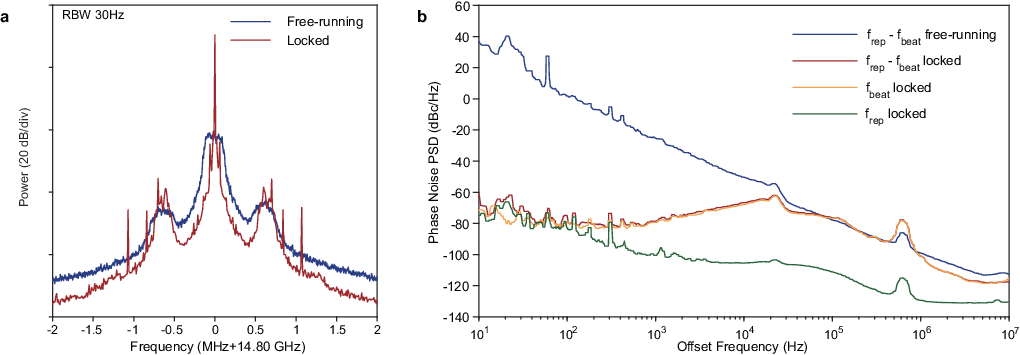}
    \caption{{\bf Full phase stabilization.} {\bf a,} Free-running and locked RF spectra of $f_{\rm rep}-f_{\rm beat}$ beatnote tone. {\bf b,} Phase noise of free-running and locked $f_{\rm rep}-f_{\rm beat}$ beatnote, compared with locked $f_{\rm beat}$ beatnote and locked repetition rate $f_{\rm rep}$.}
    \label{Ext Datafig 4}
\end{figure*}
\end{document}